%% file: entropy4.tex
\documentclass[12pt,letterpaper]{article}

\usepackage{times}
\usepackage{amssymb,amsmath}
\usepackage{epsf}
\usepackage{times,bm,mathrsfs}  
\usepackage{amsmath}
\usepackage{amssymb}
\usepackage{amsfonts}
\usepackage{mathrsfs}
\usepackage{bbm}
\usepackage{color}
\usepackage{graphicx}
\usepackage{amscd}
\usepackage{epsfig}

\definecolor{Red}{rgb}{1,0,0}
\definecolor{Blue}{rgb}{0,0,1}
\definecolor{Olive}{rgb}{0.41,0.55,0.13}
\definecolor{Green}{rgb}{0,1,0}
\definecolor{MGreen}{rgb}{0,0.8,0}
\definecolor{DGreen}{rgb}{0,0.55,0}
\definecolor{Yellow}{rgb}{1,1,0}
\definecolor{Cyan}{rgb}{0,1,1}
\definecolor{Magenta}{rgb}{1,0,1}
\definecolor{Orange}{rgb}{1,.5,0}
\definecolor{Violet}{rgb}{.5,0,.5}
\definecolor{Purple}{rgb}{.75,0,.25}
\definecolor{Brown}{rgb}{.75,.5,.25}
\definecolor{Grey}{rgb}{.5,.5,.5}
\definecolor{Black}{rgb}{0,0,0}

\def\path{{\tt path}}

\newcommand{\bcal}{\mathcal{B}}

\newcommand{\hcal}{\mathcal{H}}

\newcommand{\lcal}{\mathcal{L}}

\newcommand{\tcal}{\mathcal{T}}

\newcommand{\vcal}{\mathcal{V}}

\newcommand{\eps}{\varepsilon}

\newcommand{\ind}{\mathbbm{1}}

\newcommand{\bdm}{\begin{displaymath}}
\newcommand{\edm}{\end{displaymath}}
\newcommand{\bea}{\begin{eqnarray*}}
\newcommand{\eea}{\end{eqnarray*}}
\newcommand{\bean}{\begin{eqnarray}}
\newcommand{\eean}{\end{eqnarray}}

\newcommand{\prob}{\mathbb{P}}

\newcommand{\bfmu}{\boldsymbol{\mu}}

\newcommand{\itemname}[1]{$\mathrm{[#1]}$}
\newcommand{\bfp}{\mathbf{p}}

\newcommand{\cfn}{\mathrm{CFN}}
\newcommand{\bflambda}{\boldsymbol{\lambda}}
\newcommand{\aml}{\mathrm{AML}}
\newcommand{\qcal}{\mathcal{Q}}
\newcommand{\xx}{\mathbb{X}}
\newcommand{\hh}{\mathbb{H}}
\newtheorem{theorem}{Theorem}
\newtheorem{proposition}{Proposition}

\newtheorem{definition}{Definition}

\newtheorem{claim}{Claim}
\newenvironment{proof}{\noindent{\textbf{Proof:}}}{$\blacksquare$\vskip\belowdisplayskip}

\title{Shrinkage Effect in Ancestral Maximum Likelihood}
\author{Elchanan~Mossel\thanks{Email: mossel@stat.berkeley.edu. Depts. of Statistics and Computer Science, U.C. Berkeley. 
 Supported by an Alfred Sloan fellowship in Mathematics, by NSF grants DMS-0528488, DMS-0548249 (CAREER), 
and by DOD ONR grant N0014-07-1-05-06.}\\
Sebastien~Roch\thanks{Email: Sebastien.Roch@microsoft.com. Theory Group, Microsoft Research.}\\
Mike~Steel \thanks{Email: m.steel@math.canterbury.ac.nz. Biomathematics Research Centre, University of Canterbury,
Christchurch, New Zealand}}

\begin{document}

\maketitle
\begin{abstract}
Ancestral maximum likelihood (AML) is a method that simultaneously reconstructs a phylogenetic tree and ancestral sequences from extant data (sequences at the leaves). The tree and ancestral sequences maximize the probability of observing the given data under a Markov model of sequence evolution, in which branch lengths are also optimized but constrained to take the same value on any edge across all sequence sites. AML differs from the more usual form of maximum likelihood (ML) in phylogenetics because ML averages over all possible ancestral sequences. 
ML has long been know to be statistically consistent -- that is, it converges on the correct tree with probability approaching 1 as the sequence length grows. However, the statistical consistency of AML has not been formally determined, despite informal remarks in a literature that dates back 20 years.  In this short note we prove a general result that implies that AML is statistically inconsistent. In particular we show that AML can `shrink' short  edges in a tree, resulting in a tree that has no internal resolution as the sequence length grows.  Our results apply to any number of taxa.
\end{abstract}

\section{Introduction}
Markov models of site substitution in DNA are the basis for most methods for inferring phylogenies (evolutionary trees) from aligned sequence data.  The usual approach is maximum likelihood (ML) which seeks the tree and branch lengths that maximizes the probability of generating the observed data under a Markov process. In the simplest setting one assumes that sites evolve independently and identically, and that the extant sequences (data) label the leaves of the tree -- for background on phylogenetics and ML see \cite{fels}.
ML is computational complicated, and even the problem of finding the optimal branch lengths exactly on a fixed tree has unknown complexity. In ML one considers all possible ancestral sequences that could have existed within the tree, and averages each such `scenario' by its probability. An alternative is to simply consider a single choice of ancestral sequences that
has the highest probability -- this is a variant of ML that was introduced in 1987 by Barry and Hartigan \cite{barry1} under the name `most parsimonious likelihood', and which later was renamed {\em ancestral maximum likelihood} (AML) (see e.g. \cite{addario}).
The computational complexity of AML is slightly easier than ML, in that given the tree and either the optimal branch lengths or the optimal ancestral sequences, the other `unknown' (ancestral sequences or branch length) is readily determined (see eg. \cite{alon}). The method can be viewed as being, in some sense, intermediate between ML and a primitive cladistic method, maximum parsimony (MP), which seeks the tree and ancestral sequences that minimizes the total number of sites substitutions required to describe the data. Indeed, AML would select the same trees as MP if one further constrained AML so that each edge had the same branch length, as shown in \cite{gol}.

The recent interest in AML has sprung from computational complexity considerations.  Firstly, AML seemed to provide a promising route by which to show that the problem of reconstructing an ML tree from sequences is NP-hard \cite{addario, chor}. It turned out that the NP-hardness of ML can be established directly, without invoking AML \cite{roch}, however  the relative computational simplicity of AML over ML suggests it may provide an alternative strategy for reconstructing large trees. 

Nevertheless, it is important to know whether the desirable statistical properties of ML carry over to methods such as AML. In particular ML has long been known to be statistically consistent as a way of estimating tree topologies -- that is, as the sequence length grows, the probability that ML will reconstruct the tree that generated the sequences tends to $1$. It has also been known (since 1978) that more primitive methods, such as MP, can be statistically inconsistent \cite{felsen}.  

However the statistical consistency of AML is unclear, since the standard Wald-style conditions required to prove consistency (in particular a fixed parameter space that does not grow with the size of the data) does not apply. Thus, one may suspect that AML might be inconsistent, and indeed remarks in the literature have suggested this could be the case (see \cite{barry2}, \cite{goloboff}). However the absence of a sufficient condition to prove consistency does not constitute proof of inconsistency, and the purpose of this short note is to formally show that AML is statistically inconsistent. More precisely we show that AML tends to `shrink' short edges in a tree, and this can result in the collapse of the interior edges (and any short pendant edges) to produce a star tree.  

The results in this paper rely on probability arguments, based on expansions of the entropy function, and combinatorial properties of minimal sets of edges that separate each pair of leaves in a tree.

\subsection{Problem Statement}

\paragraph*{CFN model}
We define $[n] = \{0,\ldots,n-1\}$ and
we deal with the \emph{Cavender-Farris-Neyman (CFN) model}~\cite{Cavender:78,Farris:73,Neyman:71}. 
\begin{definition}[CFN model]
We are given a tree $T = (V,E)$ on $n$ leaves labelled $[n]$ 
and an assignment of edge probabilities
$\bfp: E \to (0,1/2)$. 
A realization of the model is obtained as follows: choose any vertex
as a root; pick a state for the root uniformly at random in $\{0,1\}$; moving
away from the root, each edge $e$ flips the state of its ancestor with probability
$p_e$. We denote by $X$ the (random) state at the leaves obtained in this manner. 
We write $X \sim \cfn(T,\bfp)$. 
\end{definition}

\paragraph*{Ancestral Maximum Likelihood}
We consider two equivalent formulations of the \emph{Ancestral Maximum Likelihood problem}.
The second version is obtained by setting
\begin{equation}\label{eq:optimalp}
p_e = \frac{d_e}{k},
\end{equation}
for all $e$ in the first version~\cite{addario}.
\begin{definition}[AML, Version 1]
The \emph{Ancestral Maximum Likelihood (AML)} problem can be stated
as follows.
Given a set of $n$ binary sequences of length $k$,
find a tree $T = (V,E)$ on $n$ leaves, an assignment $\bfp: E \to [0,1/2]$
of edge probabilities, and an assignment of sequences $\bflambda : V \to \{0,1\}^k$
to the vertices such that:
\begin{enumerate}
\item The sequences at the leaves under $\bflambda$ are exactly the sequences from $S$;
\item The quantity 
\begin{equation*}
\lcal(T, \bfp\ |\ \bflambda) = -\log_2 \left(\prod_{e \in E} p_e^{d_e} (1 - p_e)^{k - d_e}\right),
\end{equation*}
is minimized, where
\begin{equation*}
d_{u,v}
= \|\lambda_u - \lambda_v\|_1.
\end{equation*}
\end{enumerate}
\end{definition}
\begin{definition}[AML, Version 2~\cite{addario}]\label{def:version2}
The \emph{Ancestral Maximum Likelihood (AML)} problem can alternatively be stated
as follows.
Given a set of $n$ binary sequences of length $k$,
find a tree $T$ on $n$ leaves and an assignment of sequences $\bflambda : V \to \{0,1\}^k$
to the vertices such that:
\begin{enumerate}
\item The sequences at the leaves under $\bflambda$ are exactly the sequences from $S$;
\item The quantity 
\begin{equation*}
\hcal(T\ |\ \bflambda) = \sum_{e \in E} H\left(\frac{d_e}{k}\right),
\end{equation*}
is minimized, where recall that the entropy function is
\begin{equation*}
H(p) = -p \log_2 p - (1-p) \log_2 (1-p),
\end{equation*}
for $0 \leq p \leq 1$. 
\end{enumerate}
\end{definition}

\paragraph*{Consistency}
A \emph{phylogeny estimator} $\Phi = \{(\Phi_{n}^{(k)})_{n,k \geq 1}\}$ is a collection of mappings from sequences to trees,
that is,
\begin{equation*}
\Phi_n^{(k)} : \bcal_{n}^{(k)} \to \tcal_n,
\end{equation*}
where $\bcal_{n}^{(k)}$ is the set of all assignments of the form
\begin{equation*}
\bcal_n^{(k)} = \{\bfmu\ |\ \bfmu\ :\ [n] \to \{0,1\}^k\},
\end{equation*}
and $\tcal_n$ is the set of all trees on $n$ leaves labelled by $[n]$.
Let $\xx = \{X_1,X_2,\ldots\}$ with $X_j\ :\ [n] \to \{0,1\}$ for $n \geq 1$.
For all $k \geq 1$, we denote by $\bfmu = \bfmu_\xx^{(k)}$ the assignment in $\bcal_n^{(k)}$ such that
$(\mu_v)_j = (X_j)_v$ for all $v \in [n]$ and $j = 1,\ldots,k$.
\begin{definition}[Consistency]
A phylogeny estimator $\Phi$ is said to be (statistically) \emph{consistent} if for all $n$, all trees $T = (V,E) \in \tcal_n$,
and all edge probability assignments $\bfp\ :\ E \to (0,1/2)$, it holds that
\begin{equation*}
\Phi_n^{(k)}(\bfmu_\xx^{(k)}) \to T,
\end{equation*}
almost surely as $k \to +\infty$, where $\xx = \{X_1,X_2,\ldots\}$ with $X_1,X_2,\ldots$ independently generated by $\cfn(T,\bfp)$.
\end{definition}

\subsection{Main Result}

Let $\Phi_{\aml}$ be the \emph{AML phylogeny estimator} for AML Version 1, where all edges $e$ with $p_e = 0$ have been contracted
and all edges $e$ with $p_e = 1/2$ have been removed. (Break ties arbitrarily.)
\begin{theorem}[AML Is Not Consistent]\label{thm:main}
For all $n \geq 1$ and each tree $T=(V,E) \in \tcal_n$, there is a $\beta > 0$ 
and a \emph{shrinkage zone} $\qcal_T = \prod_{e \in E} I_e$ such that 
$|I_e| > \beta$ for all $e$ and  
if $\bfp \in \qcal_T$, $\Phi_\aml$ returns
a star rooted at $0$ in the limit $k \to +\infty$ on the dataset
$\xx = \{X_1,\ldots\,X_k\}$ with $X_1,\ldots,X_k$ independently generated by $\cfn(T,\bfp)$.
\end{theorem}
The phenomenon described in Theorem~\ref{thm:main} is illustrated in Fig. 1.
We note that our result does not imply the stronger statement that AML is
``positively misleading'' since we can think of the rooted star as the correct tree
$T$ where several edges are set to $p_e = 0$. Note however that the solution is
highly degenerate since the star can be obtained in this way from any tree.
In other words, in the shrinkage zone, AML provides no information about the internal
structure of the tree even with infinitely long sequences.

\begin{center} 
\begin{figure}[ht]  \label{fig1}
\resizebox{8.5cm}{!}{
\input{entropy.pstex_t} 
}
\begin{center}
\caption{The shrinkage effect: For the tree on the left, AML will reconstruct the star tree (right) from sufficiently long sequences}
\end{center}
\end{figure}
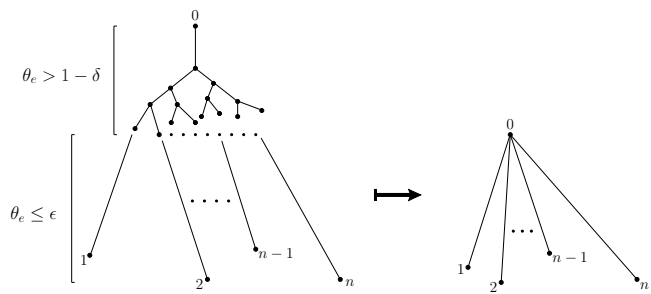
\end{center}

\subsection{Organization}

We begin with some preliminary remarks in Section~\ref{section:prelim}. The proof
of Theorem~\ref{thm:main} can be found in Section~\ref{section:proof}.

\section{Preliminaries}\label{section:prelim}

\subsection{Solution Properties}

\paragraph*{Fixed Extension} Let $T \in \tcal_n$.
For an assignment of sequences $\bfmu \in \bcal_n^{(k)}$ and $1 \leq j \leq k$,
we call $\chi\ :\ [n] \to \{0,1\}$ with $\chi_u = (\mu_u)_j$ for all $u \in [n]$
the \emph{$j$-th character} in $\bfmu$. We write $\chi \in \bfmu$ if there is $j$ such that
$\chi$ is the $j$-th character in $\bfmu$. We also denote by $\chi^\#$ the number of characters in
$\bfmu$ equal to $\chi$. An extension of a character $\chi$ is a mapping $\bar\chi\ :\ V \to \{0,1\}$
such that $\bar\chi_v = \chi_v$ for all $v \in [n]$. We denote by $\vcal(\chi)$ the set of all
extensions of $\chi$ on $T$. Let $f : \{0,1\}^{[n]} \to \{0,1\}^{V - [n]}$. The mapping then 
defines an extension for all characters simultaneously by setting $(\bar\chi_f)_v = \chi_v$ for all
$v \in [n]$ and $(\bar\chi_f)_v = f(\chi)_v$ for all $v \in V - [n]$. We show next that AML is in fact
equivalent to finding such an $f$, which can significantly reduce the size of the problem
for large $k$. For a set of $n$ binary sequences $\bfmu \in \bcal_n^{(k)}$ and a tree $T = (V,E) \in \tcal_n$,
we denote by $\bar\bfmu_f$ the extension of $\bfmu$ to $V$ by applying $f$ as above to every character in $\bfmu$. 
\begin{definition}[AML, Version 3]
Given a set of $n$ binary sequences $\bfmu \in \bcal_n^{(k)}$,
find a tree $T \in \tcal_n$ and a mapping $f : \{0,1\}^{[n]} \to \{0,1\}^{V - [n]}$ such that
the quantity 
\begin{equation*}
\hcal(T\ |\ \bar\bfmu_f) = \sum_{e \in E} H\left(\frac{d_e}{k}\right),
\end{equation*}
is minimized.
\end{definition}
\begin{proposition}[AML, Version 3]\label{proposition:aml3}
There is always a solution of AML Version 1 and 2 of the form $\bflambda = \bar\bfmu_f$ 
for some $f\ :\ \{0,1\}^{[n]} \to \{0,1\}^{V - [n]}$.
\end{proposition}
\begin{proof}
Note that
\begin{eqnarray*}
\lcal(T, \bfp\ |\ \bflambda) 
&=& -\log_2 \left(\prod_{e \in E} p_e^{d_e} (1 - p_e)^{k - d_e}\right),\\
&=& -k\sum_{e\in E} \log_2 (1 - p_e) -  \\ 
& & \sum_{j=1}^k \sum_{(u,v) \in E} \ind \{(\lambda_u)_j \neq (\lambda_v)_j\} \log_2 \frac{p_e}{1 - p_e}.
\end{eqnarray*}
For fixed $\bfp$, since $\lcal$ ``decomposes'' in $j$, 
it is always possible to take the same extension for each character appearing in $\bfmu$
without affecting optimality.
Then, we can choose the optimal $\bfp$ as in~\cite{addario} to obtain the result.
\end{proof}

\paragraph*{Limit Problem} Let $T = (V,E) \in \tcal_n$. Assume as in Theorem~\ref{thm:main} that we are given a
dataset $\xx = \{X_1,X_2,\ldots\}$ with $X_1,X_2,\ldots$~i.i.d.~$\cfn(T,\bfp)$.
Fix $f : \{0,1\}^{[n]} \to \{0,1\}^{V - [n]}$. 
Let $X \sim \cfn(T,\bfp)$ and denote by $Y = \bar X_f$ the extension of
$X$ under $f$.
Also, let $\bar\bfmu_{\xx,f}^{(k)}$ be the
extension of $\bfmu_\xx^{(k)}$ under $f$.
By the Law of Large Numbers, as $k \to +\infty$, 
the quantity $\hcal(T\ |\ \bar\bfmu_{\xx,f}^{(k)})$ converges almost surely to
\begin{equation*}
\hh_{X,T}(f) = \sum_{e\in E} H(Y_e),
\end{equation*}
where, for $e = (u,v)$, $Y_e$ is the indicator that $Y_u \neq Y_v$, and $H(Y_e)$ is the entropy of $Y_e$, that is,
\begin{equation*}
H(Y_e) = H(\prob[Y_u \neq Y_v]).
\end{equation*}
Note that, by Proposition~\ref{proposition:aml3}, even as $k\to +\infty$ there are
only a constant number of mappings $f$ to consider. We say that $f$ is $\hh_{X,T}$-optimal
if $f$ minimizes $\hh_{X,T}(f)$ over all $f:\{0,1\}^{[n]} \to \{0,1\}^{V - [n]}$.
The minimum need not be unique. 
\begin{definition}[Expected AML]
Given a random variable $X$ taking values in $\{0,1\}^{[n]}$,
find a tree $T = (V,E) \in \tcal_n$ and a mapping $f : \{0,1\}^{[n]} \to \{0,1\}^{V - [n]}$ such that
the quantity 
\begin{equation*}
\hh_{X,T}(f) = \sum_{e\in E} H(Y_e),
\end{equation*}
is minimized, where $Y = \bar X_f$.
\end{definition} 
By the previous remarks and (\ref{eq:optimalp}), to prove Theorem~\ref{thm:main}
it suffices to show:
\begin{theorem}[Optimal Assignment]\label{thm:main2}
Let $T'=(V',E') \in \tcal_n$ and let
$X \sim \cfn(T',\bfp)$. 
Then there is a $\beta > 0$ 
and a \emph{shrinkage zone} $\qcal_T = \prod_{e \in E} I_e$ such that 
$|I_e| > \beta$ for all $e$ and for all 
for all $T=(V,E) \in \tcal_n$, the unique $\hh_{X,T}$-optimal
$f:\{0,1\}^{[n]}\to \{0,1\}^{V - [n]}$ assigns to all internal
nodes of $V$ the value at leaf $0$ under all characters, that is,
\begin{equation*}
f(x) = (x_0,\ldots,x_0),
\end{equation*}
for all $x\in \{0,1\}^{[n]}$.
\end{theorem} 

\subsection{Minimal Isolating Sets}

\paragraph*{Definition} In preparation for our proof of Theorem~\ref{thm:main2}, we will need
the following notion which is studied in~\cite{MoultonSteel:04}. 
\begin{definition}[Isolating Set]
Let $T=(V,E)$ be a tree. A subset $S$ of $E$ is called an \emph{isolating 
set} for $T$ if for any two leaves $u,v$ there exists an edge 
$e \in S$ on the path connecting $u$ and $v$. 
\end{definition}
The following result is proved in~\cite{MoultonSteel:04}.
\begin{proposition}[Minimal Isolating Set]\label{lemma:nminus1}
The size of a minimal isolating
set on an $n$-leaf tree is $n-1$. 
\end{proposition}
We will also need:
\begin{proposition}[One Leaf Per Component]\label{lemma:oneleaf}
Let $T$ be a tree on $n$ leaves and let $S$ be a minimal
isolating set on $T$. Consider the forest $F$ obtained from $T$ by removing
all edges in $S$. Then, each component of $F$ contains exactly one leaf of
$T$.
\end{proposition}
\begin{proof}
If a component of $F$ contains two leaves, then these cannot be
isolated under $S$, a contradiction. 
On the other hand, if a component $T'$ of $F$ does not contain a leaf,
then every edge adjacent to $T'$ in $T$ is in fact in $S$. 
But then one can remove one of these edges without losing the isolating property of $S$,
contradicting the minimality of $S$.
\end{proof}

\paragraph*{Minimally Isolating $f$} Let $T=(V,E) \in \tcal_n$ and 
$f:\{0,1\}^{[n]}\to\{0,1\}^{V - [n]}$. We denote by $S_f \subseteq E$ the set of edges
$e = (u,v)$ such that there is $x \in \{0,1\}^{[n]}$ with $f(x)_u \neq f(x)_v$.
\begin{definition}[Minimally Isolating $f$]\label{definition:mif}
We say that $f$ is \emph{minimally isolating} for $T$ if $S_f$ is a minimal isolating
set of $T$.
\end{definition}

\subsection{Random cluster parameterization}

We will sometimes require a different (`random cluster') parameterization of the CFN model.
Let $T\in \tcal_n$ and $\bfp \in [0,1]^{E}$. (Note that we allow 
$p_e$ in $[0,1]$.) We let
\begin{equation*}
\theta_e = 1 - 2p_e,
\end{equation*}
for all $e\in E$. The main property we will use is the following well-known identity.
For two leaves $u, v$ in $T$, let $\mathrm{Path}_T(u,v)$ be the set of edges on the path between $u$ and $v$.
\begin{proposition}[Path Probability]\label{proposition:path}
Let $T = (V,E) \in \tcal_n$ and $\bfp \in [0,1]^{E}$. 
Assume $X \sim \cfn(T,\bfp)$. Let $u,v$ be two leaves of $T$. Then we have
\begin{equation*}
\prob[X_u \neq X_v] = \frac{1}{2}\left(1 - \prod_{e\in \mathrm{Path}_T(u,v)} \theta_e\right).
\end{equation*}
\end{proposition}

\section{Proof}\label{section:proof}

In this section, we prove Theorem~\ref{thm:main2} from which Theorem~\ref{thm:main} follows.
The proof has two components:
\begin{enumerate}
\item \itemname{Reduction\ to\ Minimal\ Isolating\ Sets} 
We first show that for any random variable $X \in \{0,1\}^{[n]}$
close enough to uniform and any tree $T \in \tcal_n$, the $\hh_{X,T}$-optimal
$f$'s are minimally isolating for $T$.

\item \itemname{Rooted\ Star\ is\ Optimal}
Second, we show that if $X$ above is $\cfn(T',\bfp)$ for some $T'\in \tcal_n$ with 
$p_e \approx 1/2$ if $e$ is adjacent to $\{1,\ldots, n-1\}$
and $p_e \approx 0$ otherwise, then for all
$T \in \tcal_n$ the unique $\hh_{X,T}$-optimal
$f$ assigns the value at $0$ to all internal nodes.

\end{enumerate}
Throughout, $n \geq 1$ is fixed.


\subsection{Reduction to Minimal Isolating Sets}\label{section:reduction}

We prove the following:
\begin{proposition}[Reduction to Minimal Isolating Sets]\label{proposition:entropy}
There exists $\eps > 0$ (depending on $n$) such that the following hold. Let $X$ be any random variable 
taking values in $\{0,1\}^{[n]}$ with $H(X) \geq n-\eps$ and let $T$ be any tree in $\tcal_n$. 
If $f$ is $\hh_{X,T}$-optimal, then $f$ is minimally isolating for $T$.  
\end{proposition} 
\begin{proof}
We make a series of claims.
\begin{claim}[Reduction to Uniform]
For all $\delta > 0$ there exists $\eps = \eps(\delta) > 0$ such that 
if $X$ is a $\{0,1\}^{[n]}$-random variable with
\begin{equation*}
H(X) \geq n - \eps,
\end{equation*}
and $f:\{0,1\}^{[n]}\to \{0,1\}^{V - [n]}$ then 
\begin{equation}\label{eq:ux}
| \hh_{X,T}(f) - \hh_{U,T}(f) | \leq \delta,
\end{equation} 
where $U$ is the uniform distribution on $\{0,1\}^{[n]}$. 
Therefore, it suffices to prove Proposition~\ref{proposition:entropy} for
those $f$ that are $\hh_{U,T}$-optimal.
\end{claim}
\begin{proof}
The entropy of $\{0,1\}^{[n]}$-random variables is maximized uniquely at $H(U) = n$.
The first part of the result follows by continuity of $H(X)$ and $\hh_{X,T}(f)$ in the distribution
of $X$.  

For the second part, take $\delta > 0$ small enough such that for all $f, f'$, we have
\begin{equation}\label{eq:2delta}
\hh_{U,T}(f) > \hh_{U,T}(f') \implies \hh_{U,T}(f) > \hh_{U,T}(f') + 2\delta.
\end{equation}
(Recall that there are only finitely many $f$'s for fixed $n$.) 
Take $\eps > 0$ such that the first part holds.
Then it follows that if $f$ is $\hh_{X,T}$-optimal then it must be $\hh_{U,T}$-optimal. We argue by
contradiction. Assume there are $f$, $f'$ such that
$\hh_{X,T}(f) \leq \hh_{X,T}(f')$ but $\hh_{U,T}(f) > \hh_{U,T}(f')$. By (\ref{eq:2delta}), we have
\begin{equation}
\hh_{U,T}(f) > \hh_{U,T}(f') + 2\delta,
\end{equation}
which implies $\hh_{X,T}(f) > \hh_{X,T}(f')$ by (\ref{eq:ux}), a contradiction.
\end{proof}
\begin{claim}[Minimizer]
If $f$ is $\hh_{U,T}$-optimal then $\hh_{U,T}(f) = n-1$. Moreover,
denoting $Y = \bar U_f$ we have that $\{Y_0,(Y_e)_{e\in E}\}$ are 
mutually independent.
\end{claim}
\begin{proof}

\paragraph*{Upper Bound} We first show that there is $f$ such that $\hh_{U,T}(f) \leq n-1$.
Let $S$ be a minimal isolating set for $T$. 
Define $f$ by letting $f(x)_u = f(x)_v$ for all edges $(u,v)$ not in $S$.
By Proposition~\ref{lemma:oneleaf}, this uniquely defines $f$. 
Letting $Y = \bar U_f$ it is immediate to check that 
\begin{equation*}
\hh_{U,T}(Y) = \sum_{e \in E} H(Y_e) = \sum_{e \in S} H(Y_e) \leq n-1,
\end{equation*}
by Proposition~\ref{lemma:nminus1}.

\paragraph*{Lower Bound} 
For any $f:\{0,1\}^{[n]} \to \{0,1\}^{V - [n]}$ with $Y = \bar U_f$,
we have
$$ n = H(U) = H(\{(Y_v)_{v\in [n]}\}) = H(\{Y_0,(Y_e)_{e\in E}\})$$
$$\leq  H(Y_0) + \sum_{e\in E} H(Y_e) \leq 1 + \sum_{e\in E} H(Y_e),$$
where we have used that $\{(Y_v)_{v\in [n]}\}$ and $\{Y_0,(Y_e)_{e\in E}\}$ are deterministic functions of each other.
Furthermore, the first inequality holds to equality if and only if $\{Y_0,(Y_e)_{e\in E}\}$ are mutually independent.
\end{proof}
We are ready to conclude the proof of Proposition~\ref{proposition:entropy}. 
Let $f$ be $\hh_{U,T}$-optimal with $Y = \bar U_f$. 
Let $u,v$ be any two leaves of $T$. 
We have by the previous claim that $(Y_e)_{e\in\mathrm{Path}_T(u,v)}$ are mutually independent.
Since $Y_u$ and $Y_v$ are independent uniform $\{0,1\}$ it must be that there is an edge $e \in \mathrm{Path}_T(u,v)$
with $H(Y_e) = 1$. Indeed, define $p_e = \prob[Y_e = 1]$ and $\theta_e = 1 - 2 p_e$. Then by Proposition~\ref{proposition:path}
we have
\begin{equation*}
0 = 1 - 2 \prob[Y_u \neq Y_v] = \prod_{e\in \mathrm{Path}_T(u,v)} \theta_e,
\end{equation*}
which implies that at least one $\theta_e = 0$.
Let $S'$ be the set of all edges where $H(Y_e) = 1$. Then we have shown that
$S'$ is an isolating set. Note furthermore 
that if $e \in S_f$ then $H(Y_e) \geq H(2^{-n}) > 0$. 
From $f$'s optimality we obtain
\begin{equation*}
n-1 = \hh_{U,T}(f) \geq |S'| + |S_f \setminus S'| H(2^{-n}).
\end{equation*}
Therefore we must have $S_f = S'$ and $|S'| = n-1$ which implies that $S_f$ is a minimal isolating set as needed.
\end{proof}

\subsection{The Rooted Star is Optimal}

Let $T = (V,E) \in \tcal_n$ and $S$ a minimal isolating set of $T$. Let $T^0$ be the tree
obtained from $T$ by contracting all edges not in $S$. By Proposition~\ref{lemma:oneleaf}, 
$T^0$ is an $n$-node tree 
where each node (leaf or internal) is (uniquely) labelled by a leaf of $T$.
Let $\tcal^0_n$ be all such trees on $n$ nodes.
By Proposition~\ref{proposition:entropy}, the AML phylogeny estimator is among $\tcal^0_n$.
Note that for $T \in \tcal^0_n$ the only possible extension is trivially $f = \ind$ since
there are no unlabelled internal vertices.
\begin{proposition}[Rooted Star is Optimal]\label{proposition:rootedstar}
Let $T = (V,E)\in \tcal_n $. Let $W$
be the set of leaf edges of $T$, except the edge pendant at $0$. 
Then for $\eps, \delta > 0$ sufficiently small the following holds.
Assume $X \sim \cfn(T,\bfp)$ with corresponding random cluster
parameterization satisfying $0 < \theta_e \leq \eps$ for all
$e \in W$ and $1 > \theta_e > 1-\delta$ for all $e\notin W$. 
Then, among all trees $T' \in \tcal^0_n$, the star rooted at $0$
uniquely minimizes $\hh_{X,T'}(\ind)$ for all $\delta$ sufficiently small.
\end{proposition} 
\begin{proof}
We assume that $\delta$ and $\eps$ are small enough so that they satisfy 
\begin{equation*}
(n-1)(1-\delta)^{2n-4} > n-2,
\end{equation*}
and
\begin{equation}\label{eq:eps2}
\eps^2 < (n-1)(1-\delta)^{2n-4} - (n-2).
\end{equation}

Let $T'=(V',E') \in \tcal^0_n$ and $f = \ind$ with corresponding
variables $(Y_0, \{Y_e\}_{e\in E'})$ where
$Y_0 = X_0$ and $Y_{u,v} = \ind\{X_u \neq X_v\}$. 
Let $e = (u,v)$ be an edge in $T'$. In particular, 
note that $u$ and $v$ are leaves of $T$.
Let $p_{u,v}$ be the probability that $u$ and $v$ disagree
and let $\theta_{u,v} = 1 - 2p_{u,v}$. 
We will use the following Taylor expansion of the entropy around $1/2$
\begin{equation*}
H\left(\frac{1-\tau}{2}\right)
= 1 - \left(\frac{\log_2 e}{2}\right) \tau^2 + O(\tau^4).
\end{equation*}
Note further that
\begin{equation*}
H(Y_e) = H(p_{u,v})
= H\left(\frac{1 - \theta_{u,v}}{2}\right).
\end{equation*}
As $\eps$ approaches $0$, $p_{u,v}$ goes to $1/2$.
Therefore, by Proposition~\ref{proposition:path}, up to smaller order terms we want to find
$T'=(V', E')$ in $\tcal^0_n$ that maximizes
\begin{equation*}
\Theta(T') := \sum_{e' = (u,v) \in E'} 
\prod_{e \in \mathrm{Path}_{T}(u,v)} \theta_e^2.
\end{equation*} 
If $T'$ has an edge $e'$ between two leaves neither of which is
$0$, then $e'$ makes a contribution of at most $\eps^4$ to $\Theta(T')$
since $\mathrm{Path}_{T}(u,v)$ crosses two edges in $W$. Therefore, by (\ref{eq:eps2}),
\begin{eqnarray*}
\Theta(T')
&\leq& (n-2) \eps^2 + \eps^4\\
&<& (n-1)(1-\delta)^{2n-4}\eps^2,
\end{eqnarray*}
where we have used that $T'$ has exactly $n-1$ edges and each edge $e'' \in E'$
makes a contribution of at most $\eps^2$ since $\mathrm{Path}_{T}(u,v)$ contains
at least one edge in $W$.
On the other hand, the star rooted at $0$, which we denote by $T^*$,
is the only tree in $\tcal^0_n$ which does not
include an edge between two leaves neither of which is $0$.
In that case, we get
\begin{eqnarray*}
\Theta(T^*)
&\geq& (n-1)(1-\delta)^{2(n-2)}\eps^2,  
\end{eqnarray*}
where we have used that any path between $0$ and another leaf
in $T$ contains at most $n-2$ edges not in $W$ (since $|E| \leq 2n - 3$ and $|W| = n-1$) and exactly one edge in $W$.
Taking $\eps$ small enough gives the result.
\end{proof}

\section{Concluding remarks}

It would be interesting to extend our results beyond the $2$-state case.
We note in particular that for the symmetric $r$-state model, with $r>2$, the
equivalent formulation of the AML problem given in Definition~\ref{def:version2} does not
apply. Indeed, it is easy to check that, instead, one needs to minimize
\begin{equation*}
\hcal'(T\ |\ \bflambda) = \sum_{e \in E} H\left(\frac{d_e}{k}\right) + \log_2(r-1) \sum_{e \in E} \frac{d_e}{k}.
\end{equation*}
The second term on the r.h.s.---a parsimony ``correction''---may lead to a different behavior when $r>2$.

We thank Peter Ralph for sharing his recent, independent results~\cite{ralph} regarding
the structure of the optimal solution in the $2$-state case (similarly to~\cite{alon}) as well as a number
of simulations on $4$-taxon trees.

\end{document}

%% file: entropy.pstex_t
\begin{picture}(0,0)%
\epsfig{file=entropy.pstex}%
\end{picture}%
\setlength{\unitlength}{4144sp}%
\begingroup\makeatletter\ifx\SetFigFont\undefined%
\gdef\SetFigFont#1#2#3#4#5{%
  \reset@font\fontsize{#1}{#2pt}%
  \fontfamily{#3}\fontseries{#4}\fontshape{#5}%
  \selectfont}%
\fi\endgroup%
\begin{picture}(9390,4275)(46,-3571)
\put(5041,-3481){\makebox(0,0)[lb]{\smash{\SetFigFont{14}{16.8}{\rmdefault}{\mddefault}{\updefault}\special{ps: gsave 0 0 0 setrgbcolor}$n$\special{ps: grestore}}}}
\put( 46,-2491){\makebox(0,0)[lb]{\smash{\SetFigFont{17}{20.4}{\rmdefault}{\mddefault}{\updefault}\special{ps: gsave 0 0 0 setrgbcolor}$\theta_{e}\leq\epsilon$\special{ps: grestore}}}}
\put(2746,479){\makebox(0,0)[lb]{\smash{\SetFigFont{14}{16.8}{\rmdefault}{\mddefault}{\updefault}\special{ps: gsave 0 0 0 setrgbcolor}$0$\special{ps: grestore}}}}
\put(3744,-3076){\makebox(0,0)[lb]{\smash{\SetFigFont{14}{16.8}{\rmdefault}{\mddefault}{\updefault}\special{ps: gsave 0 0 0 setrgbcolor}$n-1$\special{ps: grestore}}}}
\put(2813,-3526){\makebox(0,0)[lb]{\smash{\SetFigFont{14}{16.8}{\rmdefault}{\mddefault}{\updefault}\special{ps: gsave 0 0 0 setrgbcolor}$2$\special{ps: grestore}}}}
\put(1088,-3166){\makebox(0,0)[lb]{\smash{\SetFigFont{14}{16.8}{\rmdefault}{\mddefault}{\updefault}\special{ps: gsave 0 0 0 setrgbcolor}$1$\special{ps: grestore}}}}
\put(7441,-1141){\makebox(0,0)[lb]{\smash{\SetFigFont{14}{16.8}{\rmdefault}{\mddefault}{\updefault}\special{ps: gsave 0 0 0 setrgbcolor}$0$\special{ps: grestore}}}}
\put(6706,-3301){\makebox(0,0)[lb]{\smash{\SetFigFont{14}{16.8}{\rmdefault}{\mddefault}{\updefault}\special{ps: gsave 0 0 0 setrgbcolor}$1$\special{ps: grestore}}}}
\put(7194,-3571){\makebox(0,0)[lb]{\smash{\SetFigFont{14}{16.8}{\rmdefault}{\mddefault}{\updefault}\special{ps: gsave 0 0 0 setrgbcolor}$2$\special{ps: grestore}}}}
\put(8131,-3121){\makebox(0,0)[lb]{\smash{\SetFigFont{14}{16.8}{\rmdefault}{\mddefault}{\updefault}\special{ps: gsave 0 0 0 setrgbcolor}$n-1$\special{ps: grestore}}}}
\put(9436,-3526){\makebox(0,0)[lb]{\smash{\SetFigFont{14}{16.8}{\rmdefault}{\mddefault}{\updefault}\special{ps: gsave 0 0 0 setrgbcolor}$n$\special{ps: grestore}}}}
\put(226,-421){\makebox(0,0)[lb]{\smash{\SetFigFont{17}{20.4}{\rmdefault}{\mddefault}{\updefault}\special{ps: gsave 0 0 0 setrgbcolor}$\theta_{e}>1-\delta$\special{ps: grestore}}}}
\end{picture}